\documentclass[dvips]{acta}
\usepackage{supertabular,lscape,epsfig}
\usepackage{amssymb}
\usepackage{amsmath}
\SetPages{1}{7}     
\SetVol{69}{2019}

\begin{document}

\begin{Titlepage}

\Title { RW Tri -- its Negative Superhumps and System Parameters }

\Author {J.~~S m a k}
{N. Copernicus Astronomical Center, Polish Academy of Sciences,\\
Bartycka 18, 00-716 Warsaw, Poland\\
e-mail: jis@camk.edu.pl }

\Received{  }

\end{Titlepage}

\Abstract { Negative superhumps are detected in the light curves of RW Tri collected 
in September 1994 (Smak 1995) and November/December 1957 (Walker 1963). 
New system parameters, obtained using $K_2$ and $V_{2,rot}\sin i$ (from Poole et al. 2003) 
and $q$ (estimated from $P_{nSH}$), are 
$M_1=0.60\pm 0.20M\odot$, $M_2=0.48\pm 0.15M\odot$, $A=1.13\pm 0.09\times 10^{11}$cm 
and $i=72.5\pm 2.5$. 
}
{\it accretion, accretion disks, binaries: cataclysmic variables, 
stars: individual: RW Tri }

\section { Introduction } 

RW Tri is a nova-like cataclysmic binary. It was discovered as an eclipsing binary 
with $P_{orb}=0.23188$d by Protitch (1937) and during the next 80 years was observed 
photometrically and spectroscopically by many authors. 

The light curves of RW Tri (Smak 1995 and references therein) are peculiar. In the phase 
interval between $\phi_{orb}\sim 0.7$ and 0.9, where an orbital hump (due to the hot spot) 
should normally be present, the light curve is either nearly flat or even shows a depression 
(cf. Fig.1 in Smak 1995); similar effects were observed in the ultraviolet (Mason et al. 1997). 
This was originally misinterpreted (Smak 1995) 
as being due to unspecified "circumdisk absorption", but it is now obvious that 
it is due to the effects of stream overflow (cf. Smak 2007). 

The spectrum of RW Tri (Kaitchuck et al. 1983, Groot et al. 2004, Noebauer et al. 2010  
and references therein) is dominated by strong emission lines of hydrogen and helium. 
As noted by Groot et al. (2004), their behavior appears typical for the SW Sex 
sub-class of nova-like binaries. 
The emission lines are superpositions of the double line produced in the disk, the line 
produced in the overflowing stream and absorption effects in the stream. 
As a result, the radial velocity variations measured from different lines show different 
amplitudes and different phase shifts and are not representative for 
the orbital motion of the primary component. 

Situation improved when Poole et al. (2003), using their infrared spectra of RW Tri, 
determined $K_2$ and the rotational velocity of the secondary component and used them 
to evaluate the masses of the components (for further comments see Section 3). 

The purpose of the present paper is twofold: (1) to report on the detection of negative 
superhumps (Section 2) and (2) to redetermine system parameters (Section 3). 
 
\section { Detection of Negative Superhumps }   

The observational data which were subject to our analysis consisted of the BV light 
curves collected during twelve nights in the first half of September 1994 (Smak 1995). 
For reasons mentioned in the Introduction only parts of those light curves with 
$0.1<\phi_{orb}<0.7$ were used. 
Prior to calculating the periodograms the systematic trends with time and $\phi_{orb}$ 
have been removed using second order polynomials. 

\vskip -10truemm
\begin{figure}[htb]
\epsfysize=10.0cm 
\hspace{1.0cm}
\epsfbox{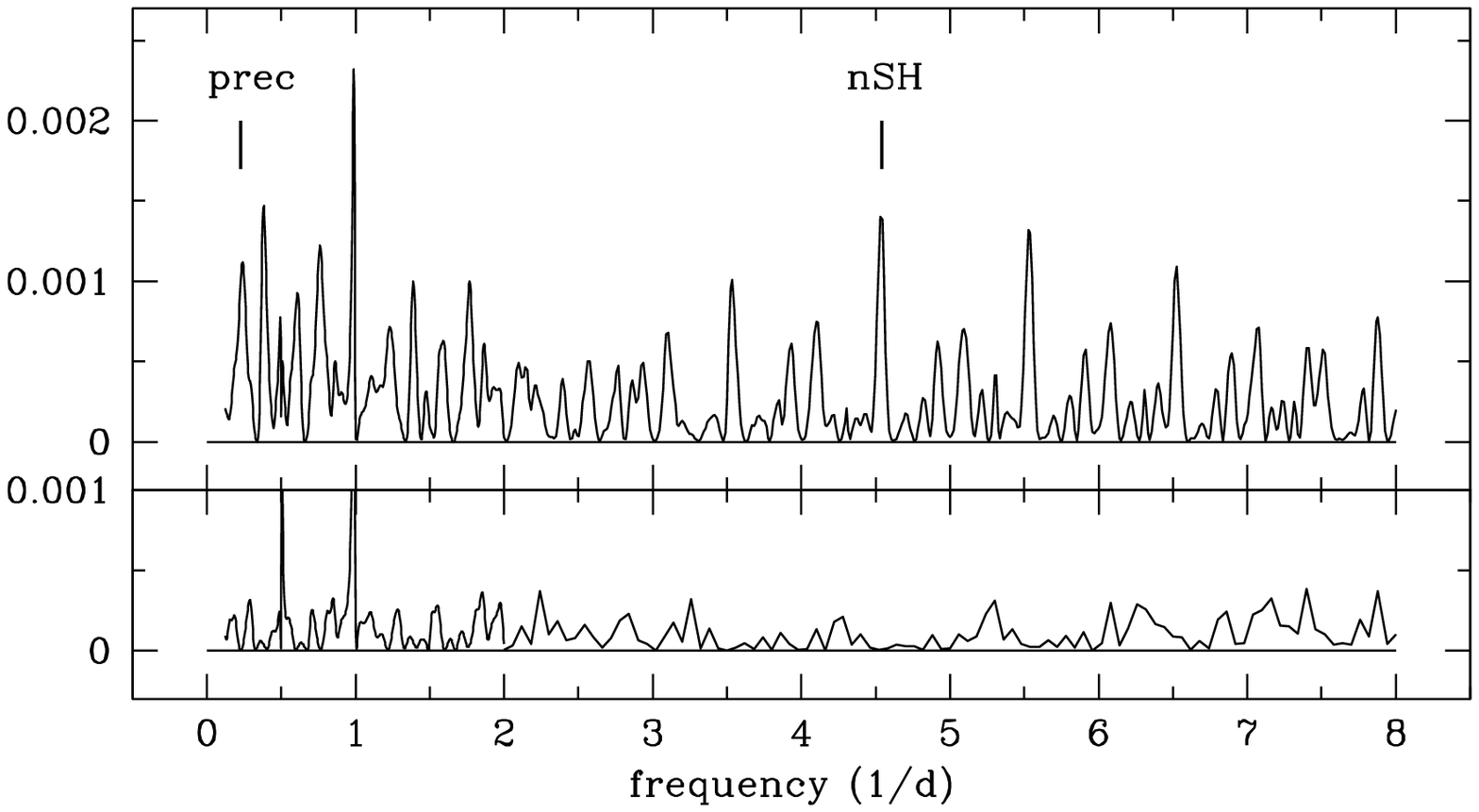} 
\vskip -40truemm
\FigCap { {\it Top:} Power spectrum of the September 1994 light curves. 
{\it Bottom:} The same after prewhitening with $P_{nSH}=0.2203$d and $P_{prec}=4.40$d. 
See text for details. 
}
\end{figure}

Fig.1 presents the periodogram calculated from B light curves (the other periodogram,  
from V light curves being practically identical). It shows several peaks: the strongest 
at $\nu=4.54\pm 0.03$d$^{-1}$ and its aliases at $\pm1$ and $\pm2$d$^{-1}$. 
We identify it as representing negative superhumps with $P_{nSH}=0.2203\pm0.0014$d 
and $\epsilon_{nSH}=0.050\pm 0.006$. 
Within the precessing, tilted disk model for negative superhumps the precession 
period is related to $P_{orb}$ and $P_{nSH}$ by 

\beq
{1\over {P_{prec}}}~=~{1\over {P_{nSH}}}~-~{1\over {P_{orb}}} . 
\eeq

\noindent
Using values of $P_{orb}$ and $P_{nSH}$ we predict: $P_{prec}=4.40$d and 
$\nu_{nSH}=0.227$d$^{-1}$. 
One of the peaks in the low frequency range of the periodogram is indeed located 
at this frequency. 

The periodogram obtained after prewhitening the data with $P_{nSH}=0.2203$d and 
$P_{prec}=4.40$d, presented in the bottom part of Fig.1, no longer shows peaks which 
were present in the original periodogram (the only exception are two spurious peaks at 
$\nu=0.5$ and 1.0$d^{-1}$). This implies that $P_{nSH}$ and $P_{prec}$ are the only 
periodicities and that their identification was correct. 

As the last step we calculate the mean negative superhump light curves and the mean 
precession light curves. They are shown in Fig.2. Their full aplitudes are 
$2A_{nSH}\approx 0.12$mag and $2A_{prec}\approx 0.10$mag. 

\vskip -30truemm
\begin{figure}[htb]
\epsfysize=12.0cm 
\hspace{0.1cm}
\epsfbox{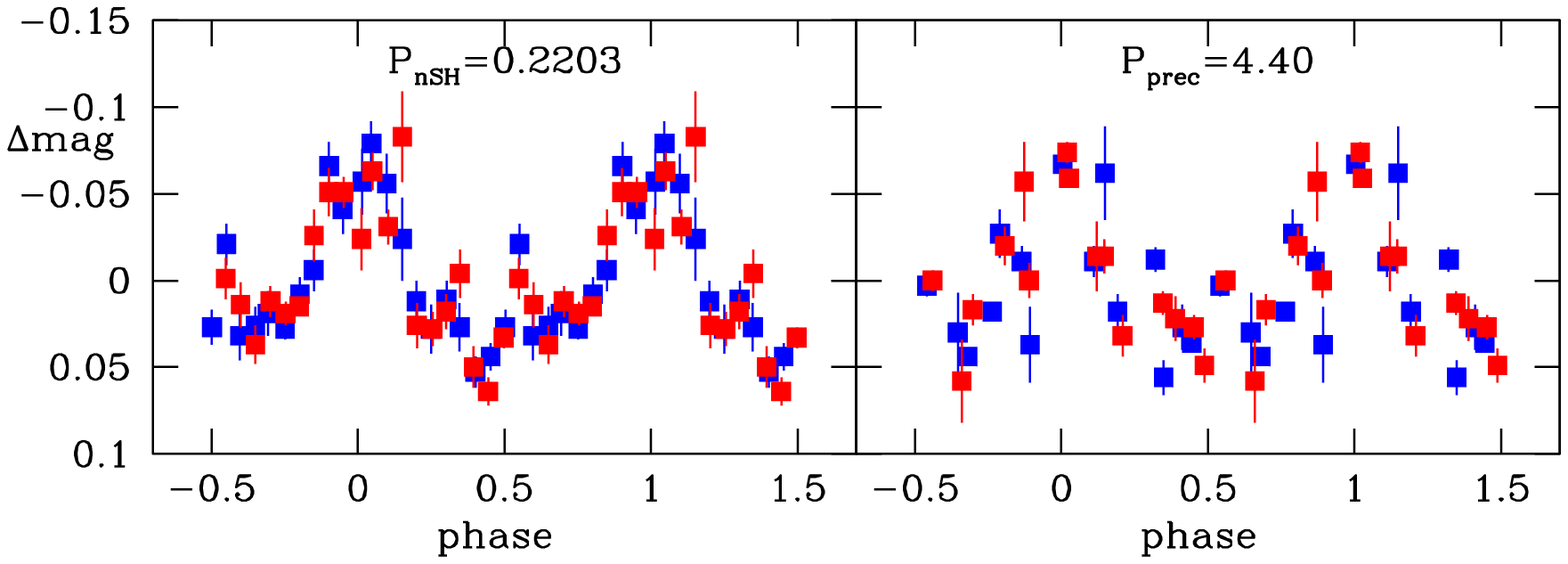} 
\vskip -45truemm
\FigCap { The mean B ({\it blue}) and V ({\it red}) light curves from September 1994 
folded with $P_{nSH}=0.2203$d ({\it left}) and $P_{prec}=4.40$d ({\it right}). 
Two cycles are shown for clarity. 
}
\end{figure}

An attempt was also made to analyze light curves collected in the fall of 1957 
by Walker (1963). Regretfully, RW Tri showed at that time large variations of its mean 
brightness. Besides several light curves covered only the eclipse. 
In fact, only four yellow light curves, collected between November 27 and December 2, 1957, 
could be used for the present purpose. 
The mean brightness of the star declined during those nights by about 0.2 mag. 
Under those circumstances searching for a signal corresponding to $P_{prec}$ 
turned out to be practically impossible. 

\vskip -5truemm
\begin{figure}[htb]
\epsfysize=10.0cm 
\hspace{1.5cm}
\epsfbox{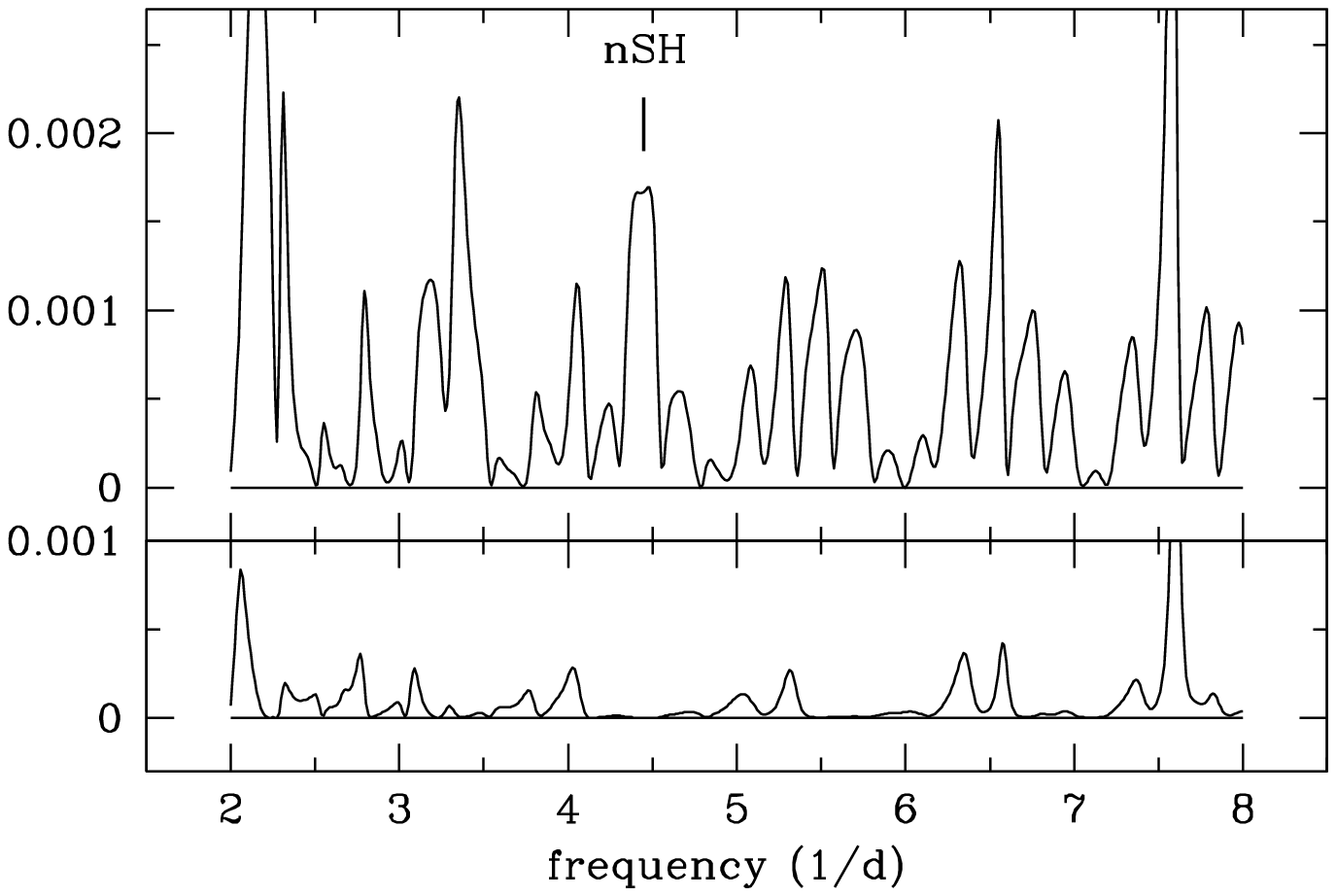} 
\vskip -40truemm
\FigCap { {\it Top:} Power spectrum of the November/December 1957 light curves. 
{\it Bottom:} The same after prewhitening with $P_{nSH}=0.2235$d. 
}
\end{figure}

The periodogram presented in Fig.3 shows several peaks, one of them at 
$\nu_{nSH}=4.50\pm 0.10$d$^{-1}$, or $P_{nSH}=0.2235\pm 0.0050$d, which is -- within 
errors -- consistent with values determined earlier. 
The periodogram obtained after prewhitening the data with this period is shown  
in the bottom part of Fig.3 and the mean light curve -- in Fig.4. 
Those results are sufficient to conclude that the negative superhumps were also present 
in November/December 1957.

\vskip -30truemm
\begin{figure}[htb]
\epsfysize=12.0cm 
\hspace{0.5cm}
\epsfbox{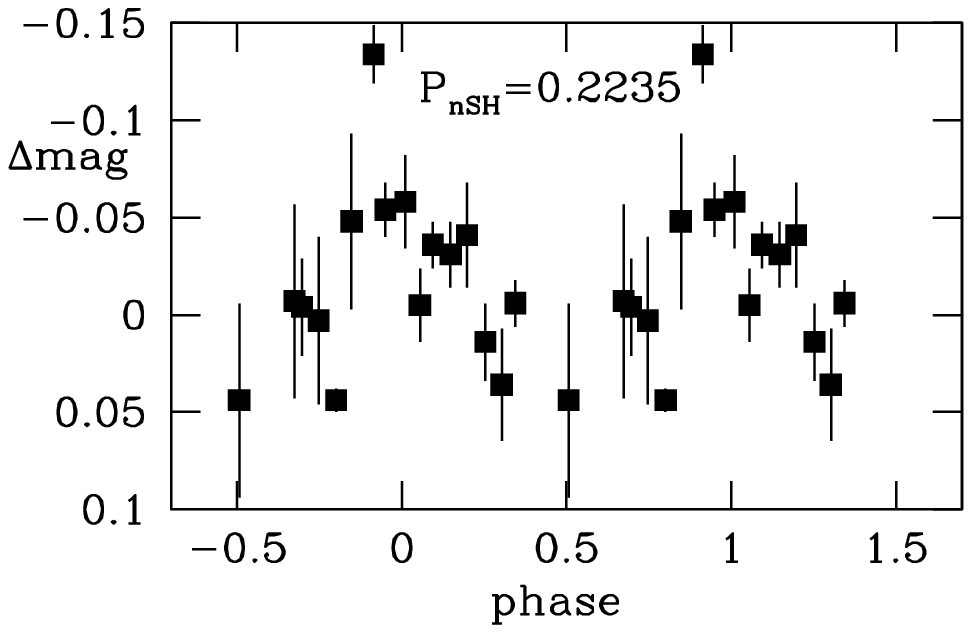} 
\vskip -45truemm
\FigCap {The mean light curve from November/December 1957 folded with $P_{nSH}=0.2235$d. 
Two cycles are shown for clarity.  
}
\end{figure}

\section { System Parameters } 

Poole et al. (2003) determined $K_2=221\pm 29$km/s and the rotational velocity of 
the secondary component $V_{2,rot}\sin i=120\pm 20$km/s. 
The value of $K_2$ corrected by them for the effect of irradiation was 
$K_{2,corr}=178$km/s, and using it together with $V_{2,rot}\sin i$ they found: 
$M_1=0.4M\odot$ and $M_2=0.3M\odot$ (with no attempt to evaluate their errors).   
Prior to that step they presented convincing evidence showing that correcting 
$K_2$ for the irradiation effect is necessary. It is therefore unclear why they 
repeated their calculations using the observed, {\it uncorrected} value of $K_2$.   

Turning to the results obtained by them with $K_{2,corr}$ it turns out that 
they were wrong. The correction to $K_2$ for the irradiation effect depends on 
the mass-ratio (see Eqs 3 and 4 in Poole et al. 2003, or Eq.4 below). 
The correction $\Delta K_2=-43$ km/s adopted by Poole et al. (2003) resulted from 
the arbitrarily assumed $q\approx 1.35$, in an obvious contradiction to the evidence 
presented in their Figs.9 and 10, to the value $q=0.85$ required by the ratio 
$K_{2,corr}/V_{2,rot}\sin i$, and to their final $q=0.3/0.4=0.75$. 
With the correct value of $q$ the correction $\Delta K_2$ would be much smaller 
(see Section 4, below) and the resulting $K_{2,corr}$ significantly larger. 

Problems described above are not the only reason for another attempt to determine 
system parameters. 
We now have -- in addition to $K_2$ and $V_{2,rot}\sin i$ -- a third constraint based 
on the period of negative superhumps (see below) which permits a more reliable 
evaluation of the mass ratio. We proceed as follows. 

To begin with, we have the standard formulae for the masses of the two components: 

\beq
M_1~=~{P_{orb}\over{2\pi}}~{{(1+q)^2}\over G}~K_{2,corr}^3~cosec^3 i~ ,
\eeq

\noindent
and

\beq
M_2~=~{P_{orb}\over{2\pi}}~{{(1+q)^2q}\over G}~K_{2,corr}^3~cosec^3 i~ ,
\eeq

\noindent
where the value of $K_2$ corrected for effects of irradiation is given by 
(cf. Poole et al. 2003, Section 4.2)

\beq
K_{2,corr}~=~{ {K_2}\over {1~-~{4\over {5\pi}}~(1+q)~r_{Roche,2} } }~ .
\eeq

\noindent
Those three equations, together with the $i=i(M_1)$ relation (Smak 1995, Fig.3), 
define the $M_2=f(M_1)$, or $M_1=f(M_2)$ relation, shown -- together with its errors 
-- with red lines in Fig.5. 

Turning to our second constraint resulting from $V_{2,rot}\sin i$ and noting that 
(cf. Poole et al. 2003, Eq.9)

\beq
V_{2,rot}\sin i~=~(1~+~q)~r_{Roche,2}~K_{2,corr}~ ,
\eeq

\noindent
we can replace Eqs.(2) and (3) with 

\beq
M_1~=~{P_{orb}\over{2\pi}}~{1\over G}~{1\over{1+q}}~{1\over {r_{Roche,2}^3}}~(V_{2,rot}\sin i)^3~cosec^3 i~ , 
\eeq

\noindent
and

\beq
M_2~=~{P_{orb}\over{2\pi}}~{1\over G}~{q\over{1+q}}~{1\over {r_{Roche,2}^3}}~(V_{2,rot}\sin i)^3~cosec^3 i~ . 
\eeq

\noindent
Those equations, together with the $i=i(M_1)$ relation, define another 
$M_2=f(M_1)$ relation, shown -- together with its errors -- with green lines in Fig.5. 

The third constraint comes from the period of negative superhumps, which -- according  
to the precessing, tilted disk model (e.g. Osaki and Kato 2013) -- is related 
to the mass-ratio and the radius of the disk by 

\beq
{3\over7}~{q\over {(1+q)^{1/2}}}~r_d^{3/2}~=~{\epsilon_{nSH}\over{1+\epsilon_{nSH}}}~ .  
\eeq

\noindent
In systems with stationary accretion the radius of the disk is controlled by tidal effects 
and is effectively equal to the so-called tidal radius $r_{tid}$, which is slightly smaller 
than the mean radius of the Roche lobe around the primary component.   
Using results obtained by Paczy{\'n}ski (1977) and Ichikawa and Osaki (1994) we adopt

\beq
r_d~=~r_{tid}~\approx~0.86~r_{Roche,1}~ .
\eeq

\noindent
Solving Eq.(8) with $\epsilon_{nSH}=0.050\pm 0.006$ (Section 2) we find 
$q=0.85\pm 0.25$, shown with blue lines in Fig.5. 

Results, presented in Fig.5, show that the three lines, representing three different 
constraints discussed above, intersect in a very small area. Therefore, in spite 
of large uncertainties (dotted lines), the masses of the components can be reliably determined.  
We adopt: $M_1=0.60\pm 0.20M\odot$ and $M_2=0.48\pm 0.15M\odot$, the errors being defined 
primarily by uncertainties in $q$. 
With those masses we get: $A=1.13\pm 0.09\times 10^{11}$cm and -- 
using the $i=i(M_1)$ relation (Smak 1995, Fig.3) -- the orbital inclination $i=72.5\pm 2.5$. 

\vskip -10truemm
\begin{figure}[htb]
\epsfysize=8.0cm 
\hspace{2.0cm}
\epsfbox{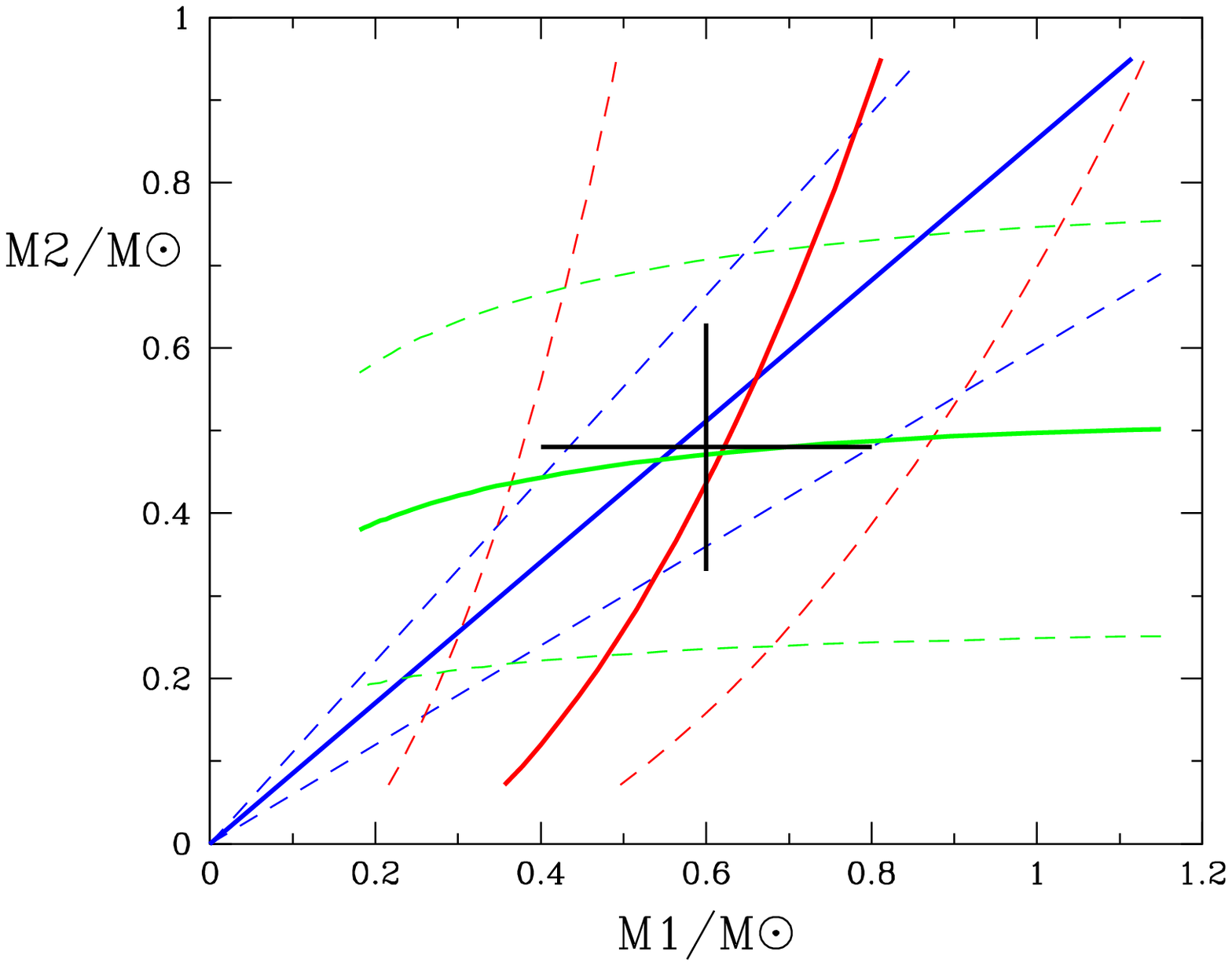} 
\vskip -15truemm
\FigCap { Masses of the components. Lines of different colors represent constraints resulting 
from $K_{2,corr}$ ({\it red}), $V_{2,rot}\sin i)$ ({\it green}) and $\epsilon_{nSH}$ ({\it blue}). 
The black cross represents the adopted values of $M_1$ and $M_2$. See text for details. 
}
\end{figure}

\section { Discussion } 

Results presented above can be supplemented with the following remarks: 

{\parskip=0truept { 
(1) The value $\epsilon_{nSH}=0.050$ for RW Tri is similar to those in two other 
longest period superhumpers: UX UMa, with $P_{orb}=0.1967$d and $\epsilon_{nSH}=0.051$ 
(de Miguel et al. 2016) and TV Col with $P_{orb}=0.2286$d and $\epsilon_{nSH}=0.054$ 
(Retter et al. 2003). 

(2) RW Tri becomes the new record-holder: it has the longest orbital period 
$P_{orb}=0.23188$d among all superhumpers; on the other hand, however, the period of 
its negative superhumps $P_{nSH}=0.2203$d is only the second longest, after that of 
the common superhumps in TV Col: $P_{SH}=0.2639$d (Retter et al. 2003). 

(3) The amplitude of the mean precession light curve can be used to evaluate (cf. Smak 2009) 
the disk tilt with respect to the orbital plane. Using $A_{prec}\approx 0.05$mag and 
$i=72.5$ we get $\delta\approx 1^o$. 

(4) System parameters obtained above can be used to predict: $K_1=158$km/s  
and $K_2=197$km/s. It turns out that the correction to $K_2$ for irradiations effects was 
$\Delta K_2=221-197=24$km/s, i.e. factor of 2 smaller than that used by Poole et al. (2003). 

(5) The mean radius of the secondary components comes out as $R2=0.59\pm 0.09R\odot$, 
i.e. larger but -- within errors -- comparable to $R2=0.54R\odot$, resulting from the 
mean $R2-M2$ relation for main sequence secondaries (Knigge et al. 2011). 
}}

\begin {references} 

\refitem {de Miguel, E., Patterson, J., Cejudo, D., Ulowetz, J., Jones, J.L., Boardman, J., 
          Barrett, D., Koff., R., Stein, W., Campbell, T., Vanmunster, T., Menzies, K., 
          Slauson, D., Goff, W., Roberts, G., Morelle, E., Dvorak, S., Hambsch, F.-J., 
          Starkey, D., Collins, D., Costello, M., Cook, M.J., Oksanen, A., Lemay, D., 
          Cook, L.M., Ogmen, Y., Richmond, M., Kemp, J.} {2016} {\MNRAS} {457} {1447}     

\refitem {Groot, P.J., Rutten, R.G.M., van Paradijs, J.} {2004} {\AA} {417} {283} 

\refitem {Ichikawa S., Osaki, Y.} {1994} {\it Publ.Astr.Soc.Japan} {46} {621} 

\refitem {Kaitchuck, R.H., Honeycutt, R.K., Schlegel, E.M.} {1983} {\ApJ} {267} {239} 

\refitem {Knigge, C., Baraffe, I., Patterson, J.} {2011} {\ApJ Suppl.} {194} {28}

\refitem {Mason, K.O., Drew, J.E., Knigge, C.} {1997} {\MNRAS} {290} {L23} 

\refitem {Noebauer, U.M., Long, K.S., Sim, S.A., Knigge, C.} {2010} {\ApJ} {719} {1932} 

\refitem {Osaki, Y., Kato, T.} {2013} {\it Publ.Astr.Soc.Japan} {65} {95}

\refitem {Paczy{\'n}ski, B.} {1977} {\ApJ} {216} {822}

\refitem {Poole, T., Mason, K.O., Ramsay, G., Drew, J.E., Smith, R.C.} 
      {2003} {\MNRAS} {340} {499}

\refitem {Protitch, M.} {1937} {Bull.Astr.Obs.Belgrade} {9-10} {38}

\refitem {Retter, A., Hellier, C., Augusteijn, T., Naylor, T., Bedding, T.R., 
          Bembrick, C., McCormick, J., Velthuis, F.} {2003} {\MNRAS} {340} {679}

\refitem {Smak, J.} {1995} {\Acta} {45} {259}

\refitem {Smak, J.} {2007} {\Acta} {57} {87}

\refitem {Smak, J.} {2009} {\Acta} {59} {419}

\refitem {Walker, M.F.} {1963} {\ApJ} {137} {485}

\end {references}

\end{document}